\documentclass[doublecol]{epl2}
\usepackage{xcolor,soul}

\title{Neutron powder diffraction study on the iron-based nitride superconductor ThFeAsN}
\shorttitle{Neutron powder diffraction of ThFeAsN} %Insert here a short version of the title if it exceeds 70 characters

\author{Huican Mao\inst{1,2}, Cao Wang\inst{3}, HELEN E. Maynard-Casely\inst{4}, Qingzhen Huang\inst{5}, Zhicheng Wang\inst{6}, Guanghan Cao\inst{6,7}, Shiliang Li\inst{1,2,8} \and Huiqian Luo\inst{1}$^{\dag}$}
\shortauthor{H. Mao \etal}

\institute{
  \inst{1}Beijing National Laboratory for Condensed Matter Physics, Institute of Physics,
  Chinese Academy of Sciences, Beijing 100190, China\\
  \inst{2}University of Chinese Academy of Sciences, Beijing 100049, China\\
  \inst{3}Department of Physics, Shandong University of Technology, Zibo 255049, China\\
  \inst{4}Australian Centre for Neutron Scattering, Australian Nuclear Science and Technology Organisation, Lucas Heights NSW-2232, Australia\\
   \inst{5}NIST Center for Neutron Research, National Institute of Standards and Technology,
Gaithersburg, Maryland 20899-6102, USA\\
  \inst{6}Department of Physics and State Key Lab of Silicon Materials, Zhejiang University,
Hangzhou 310027, China\\
  \inst{7}Collaborative Innovation Centre of Advanced Microstructures, Nanjing 210093, China\\
  \inst{8}Collaborative Innovation Center of Quantum Matter, Beijing 100190, China\\

}

\abstract{
We report neutron diffraction and transport results on the newly discovered superconducting nitride ThFeAsN with $T_c=$ 30 K. No magnetic transition, but a weak structural distortion around 160 K, is observed cooling from 300 K to 6 K. Analysis on the resistivity, Hall transport and crystal structure suggests this material behaves as an electron optimally doped pnictide superconductors due to extra electrons from nitrogen deficiency or oxygen occupancy at the nitrogen site, which together with the low arsenic height may enhance the electron itinerancy and reduce the electron correlations, thus suppress the static magnetic order.}

\begin{document}

\maketitle

\section{Introduction}

 Unconventional superconductivity in iron pnictides or chalcogenides has been intensively investigated since the ZrCuSiAs-type crystalline LaFeAsO$_{1-x}$F$_x$ (1111 family) with transition temperature $T_c=26$ K  was discovered in 2008 \cite{Hosono1}. Usually, the iron-based superconductivity emerges from the proximity to a three-dimensional antiferromagnetism \cite{Dai1}, for example, LaFeAsO$_{1-x}$F$_x$ \cite{CruzC,HuangQ}, BaFe$_{2-x}$Co$_{x}$As$_2$(122 family) \cite{Shibauchi}, NaFe$_{1-x}$Co$_x$As(111 family) \cite{Parker}, FeTe$_{1-x}$Se$_x$(11 family) \cite{LiuTJ,KatayamaN1} and Ca$_{1-x}$La$_{x}$FeAs$_{2}$ (112 family) \cite{KatayamaN2,JiangS} etc. In some special cases, the conductivity is very sensitive to the ion deficiency, such as LaFeAsO$_{1-\delta}$ \cite{RenZA}, Li$_{1-\delta}$FeAs \cite{WangXC} and K$_{0.8}$Fe$_{2-\delta}$Se$_2$ \cite{GuoJG}, or the stoichiometric composition is naturally superconducting, such as Sr$_2$VO$_3$FeAs (21311 family) \cite{ZhuX1}, Ca$_{10}$(Fe$_3$Pt$_8$)(Fe$_2$As$_2$)$_5$ (10-3-8 family) \cite{NiN}, KFe$_2$As$_2$ \cite{ChenH}, RbEuFe$_4$As$_4$ (1144 family) \cite{LiuY}, FeSe \cite{HsuF}, etc. Even so, in most of above families, spin fluctuations persist and intimately interplay with superconductivity, while in most parent compounds, magnetic phase transition always occurs beneath the symmetry-breaking structural transition at low temperature except for the FeSe system \cite{WangQS}.

Specifically for the 1111 family, superconductivity can be induced among the antiferromagnetically ordered oxide (e.g. LaFeAsO) \cite{Hosono1}, fluoride (e.g. CaFeAsF) \cite{Hosono2,ZhuX2} and hydride (e.g. LaFeAsO$_{1-x}$H$_x$) \cite{Hosono3,Hosono4} under the chemical substitution on any atomic site. Recently, the first nitride iron pnictide superconductor ThFeAsN, containing layers with nominal compositions [Th$_2$N$_2$] and [Fe$_2$As$_2$] (Fig.1(a)), has been discovered, with $T_c=$ 30 K for nominally undoped compound \cite{Wang1}. Although the first-principle calculation of ThFeAsN indicates the lowest energy magnetic ground state is the stripe-type antiferromagnetic state \cite{Guang1,Singh1}, the normal state resistivity shows no obvious anomaly but a metallic behavior. In principle, the N-N bond covalency may lower the effective nitrogen valence and lead to an internal charge transfer. Such self-doping effect may be responsible for the superconductivity and suppress the magnetic order completely, since any further electron doping via substituting N with O or hole doping via substituting Th with Y only suppress the supperconducting $T_c$ \cite{Wang1}. Indeed, the $^{57}$Fe M\"ossbauer spectroscopy study on the polycrystalline samples suggests no magnetically ordered moment on iron site down to 2 K \cite{Albedah1}. To finally clarify the absence of magnetic order and classify the superconductivity with other compounds, neutron powder diffraction experiments are highly desired for this new material.

\begin{figure}
\onefigure[width=2.6in]{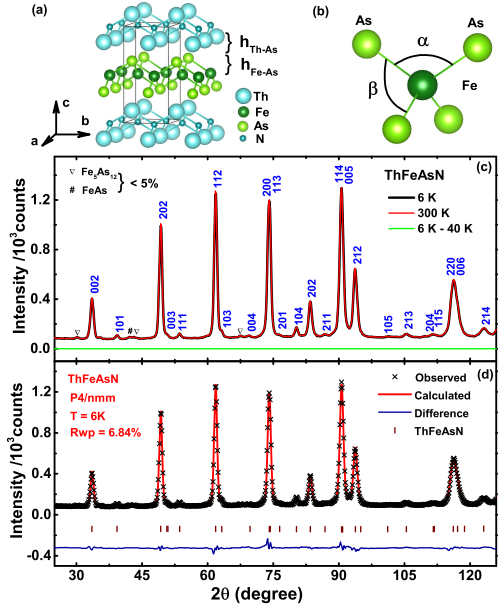}
\caption{(a) The crystal structure and (b) FeAs$_4$ tetrahedron of ThFeAsN. Where $h_{Th-As}$ and $h_{Fe-As}$ are arsenic height to Th layer and Fe layer, respectively. $\alpha$ and $\beta$ indicate the two different As-Fe-As bond angles\cite{Wang1}.(c) Normalized neutron powder diffraction patterns at 6 K and 300 K and the difference between 6 K and 40 K for ThFeAsN. A few peaks of Fe$_5$As$_{12}$ and FeAs impurity phase are marked. (d) Rietveld refinement results with tetragonal phase ${P4/nmm}$ space group at 6 K.}
\label{fig.1}
\end{figure}

\section{Experiments}
The polycrystalline samples with $T_c=$ 30 K were synthesized by the solid state reaction method as described elsewhere [6]. To make sure homogeneous scattering background, 2 grams of powder samples were ground and sealed in a vanadium can. Neutron powder diffraction experiments were carried out on WOMBAT high-intensity diffractometer at Australian Centre for Neutron Scattering, Australian Nuclear Science and Technology Organisation. The wavelength of neutron was selected to be $\lambda$ = 2.41 \AA. The scattering data was collected at 6 K (4 hrs), 40 K (4 hrs) and other temperatures up to 300 K (1 hr for each) by covering the scattering angle $2\theta$ range 15 - 136 degrees. All these diffraction patterns were refined with Rietveld method within the program FullProf\cite{Juan1}, and the temperature dependence of structure parameters, such as lattice constant, full-width-at-half-maximum (FWHM) of (112) peak, As-Fe-As bond angles, bond length and the ionic (Th, Fe) height from As-layer, were obtained by assuming 100\% occupancy of ThFeAsN. Temperature dependent resistivity from 2 K to 300 K was measured by standard 4-probe method, and the Hall coefficient($R_H$) was measured by the transverse resistance under sweeping magnetic fields from -6 T to +6 T over the temperature range 50 K - 300 K on a \emph{Quantum Design} Physical Property Measurement System (PPMS).

\begin{figure}
\onefigure[width=3.2in]{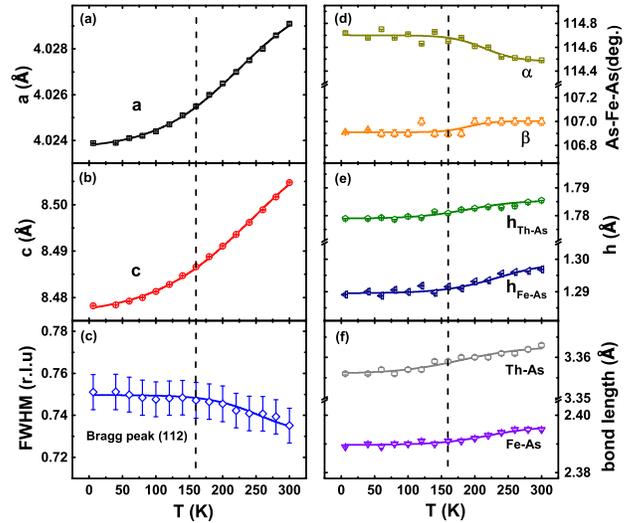}
\caption{Temperature dependence of (a) lattice constant a  (b) lattice constant c (c) FWHM of (112) Bragg peak (d) As-Fe-As bond angles $\alpha$ and $\beta$  (e) anion (Th, Fe) height from As-layer (f) bond length of Th-As and Fe-As in ThFeAsN.}
\label{fig.2}
\end{figure}

\begin{table}
\caption{Crystallographic data of ThFeAsN at 6 K.}
\label{tab.1}
\begin{centering}
{
\begin{tabular}{lccr}
\hline
space group & P4/$nmm$     & R$wp$(\%)    & 6.84(3) \\
$a$(\AA)       & 4.0414(1)      &$h_{Th-As}$(\AA)  & 1.7858(1)\\
$c$(\AA)       & 8.5152(1)    & $h_{Fe-As}$(\AA)   &1.2964(1)\\
$\alpha$$_{Fe-As-Fe}$ &$114.65^{\circ}$  &$d_{Th-As}$(\AA) &3.3700(3)\\
$\beta$$_{Fe-As-Fe}$ &$106.95^{\circ}$  &$d_{Fe-As}$(\AA) &2.4010(3)\\
\end{tabular}
}

{
\begin{tabular}{lccccr}
\hline
atom     &Wyckoff    &x   &y   &z   &U$_{iso}$\\
Th       &2c &0.25 &0.25 &0.1380(3) &0.3768\\
Fe       &2b &0.75 &0.25 &0.5 &0.0934\\
As       &2c &0.25 &0.25 &0.6522(5) &0.0884\\
N        &2a &0.75 &0.25 &0 &0.7005\\
\hline
\end{tabular}
}

\end{centering}
\end{table}

\section{Result and discussion}

\begin{figure}
\onefigure[width=2in]{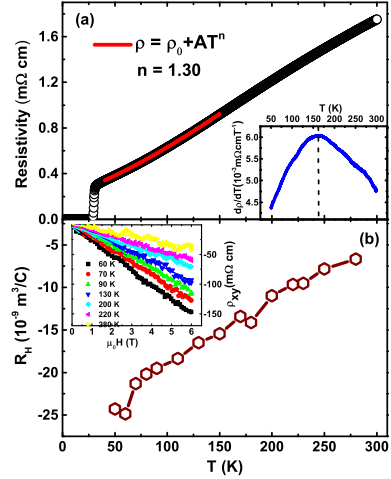}
\caption{(a) Temperature dependence of resistivity at zero magnetic field  (b) Hall coefficient R$_H$ for the ThFeAsN polycrtstalline sample. Insert (a) shows the derivative d$R$/d$T$ as function of temperature. Insert (b) shows  the magnetic field dependence of Hall resistivity at 60 K, 70 K, 90 K, 130 K, 200 K, 220 K and 280 K.}
\label{fig.3}
\end{figure}

The raw data of neutron diffraction patterns are presented in Fig.1. To quantitatively compare the neutron diffraction peaks, we have normalized the maximum neutron counts at 300 K to be the same as 6 K and the difference between 6 K and 40 K is obtained by direct subtraction from the raw data after normalized by monitor counts (Fig.1(c)). Most of reflections can be indexed by a tetragonal phase in ZrCuSiAs-type structure with the space group $P4/nmm$ ($a$=$b$=4.0414 \AA, $c$=8.5152 \AA), except for very small amount Fe$_5$As$_{12}$ and FeAs impurity phases ($<5$ \%), which are not detected in X-ray diffraction experiment \cite{Wang1,Albedah1}. Fig.1(d) shows the Rietveld refinement of the neutron diffraction patterns at 6 K by assuming 100\% occupancy of ThFeAsN. The parameters for the quality of this fitting are: profile factor Rp = 6.47(2)\%, weighted profile factor Rwp = 6.84(3)\%, and reduced $\chi$-square $\chi$$^2$ = 2.93(1). All crystallographic parameters listed in Table.1 are mostly consistent with previous X-ray diffraction results \cite{Wang1}. Since there is no difference between 6 K and 40 K data sets and all reflections are identified arising from the nuclear structure from 6 K up to 300 K, we thus conclude that there is no magnetic order in ThFeAsN, consistent with the M\"ossbauer spectroscopy results \cite{Albedah1}.

 We also performed Rietveld refinements with the FullProf program considering other possibilities. The best fitting results are in three cases: fully occupied compound (ThFeAsN), ($2.7 \pm 0.8$) \% N deficiency (ThFeAsN$_{0.97}$) or ($7 \pm 2$)\%  O occupancy at N site (ThFeAsN$_{0.93}$O$_{0.07}$), among which there is no remarkable difference of the reliability of factors between ThFeAsN. For ThFeAsN$_{0.97}$, the factors are Rp = 6.45(2)\%, Rwp = 6.82(2)\%, and $\chi$$^2$ = 2.92(2), and for ThFeAsN$_{0.93}$O$_{0.07}$ are Rp = 6.45(1)\%, Rwp = 6.82(2)\%, and $\chi$$^2$ = 2.93(1). Other cases such as Th or As deficiency give much worse factors, even failure of the fitting. Although the N deficiency or O occupancy as well as the exact proportions of them cannot be distinguished precisely within the data quality and instrument resolution, the N deficiency is more possibly indicated by the refinements due to larger neutron cross section of nitrogen.

To search for a possible structural transition in ThFeAsN, we have collected diffraction patterns over temperature range 300 K to 6 K. All of them can be well refined with the tetragonal phase of ThFeAsN same as the data at room temperature. Therefore, there is no tetragonal-to-orthorhombic transition in lattice. Fig.2 summarizes the temperature dependence of parameters from Rietveld analysis: the lattice constants ($a$, $c$), ionic (Th, Fe) height from As-layer ($h_{Th-As}$, $h_{Fe-As}$), the bond angle As-Fe-As ($\alpha$, $\beta$) and the bond length of Fe-As and Th-As. The data sets are nearly unchanged if considering N deficiency or O occupancy. Although the lattice parameters continuously increase with temperature due to thermal expansion, there is a broad kink existing around 160 K for other parameters, similar feature is also found in the peak width of (112). This can be explained by a weak distortion of the FeAs$_4$ tetrahedron (Fig.1(b)), which may not strong enough to induce a structural transition similar to some optimally doped iron pnictides \cite{LuX,HuD1}. Further experiments on the single crystal will be much helpful to clarify this issue.

\begin{figure}
\onefigure[width=2in]{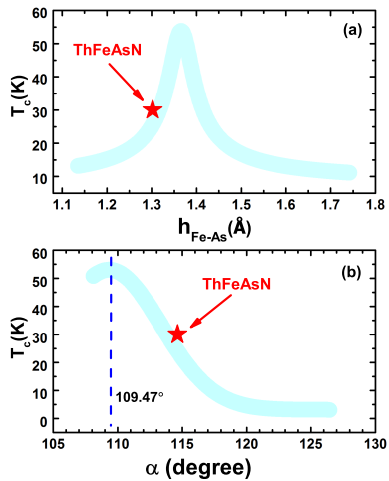}
\caption{(a) Anion height dependence of optimal $T_c$ for Fe-based superconductors. (b) optimal $T_c$ vs bond angle $\alpha$. The light cyan belts are from other optimally doped systems \cite{Chen2}, and the red stars indicate our ThFeAsN compound, respectively.}
\label{fig.4}
\end{figure}

In order to get more insight into the transport properties at normal state, we have carried out temperature dependence of resistivity and Hall effect measurements on the polycrystalline ThFeAsN, as shown in Fig.3. The resistivity data shows a metallic behavior above $T_c$ without any clear anomaly expected for a magnetic or structural transition. It can be fitted by an empirical power law $\rho$=$\rho$$_0$+$AT^n$ in a wide temperature range (40 K- 150 K) with the exponent $n$ approximate to 1.30, suggesting possible non-Fermi-liquid behaviors governed by quantum fluctuations similar to systems around the optimal doping level \cite{Shibauchi,Onari1,Kasahara1,Luo1}. Interestingly, the derivative d$R$/d$T$ (insert of Fig.3(a)) also shows a broad hump around 160 K, corresponding to the lattice distortion found in Fig.2. The Hall coefficient($R_H$), determined by the slope of field dependence of Hall resistivity, is always negative and increases with temperature. By comparing with (La, Sm)FeAsO$_{1-x}$F$_x$ \cite{Hosono5,Chen1}, the magnitude of $R_H$ in ThFeAsN suggests it is already doped by electrons, which are probably introduced by the N deficiency or O occupancy shown in the powder diffraction refinement, or the reduced valence of nitrogen as discussed before.

By summarizing the literature that has noted arsenic height $h_{Fe-As}$, As-Fe-As bond angle $\alpha$ and optimized $T_c$ for each system of iron-based superconductors, it is found that there is a close relationship between the local structure of FeAs$_4$ tetrahedron and superconducting $T_c$, as shown by the light blue belt in Fig.4 \cite{Mizuguchi1,Okabe1,Chen2}. Obviously, the maximum $T_c$ was achieved when the FeAs$_4$ tetrahedron is perfectly regular, with the bond angle of 109.47 degree \cite{Shirage1}. Interestingly, the data obtained from ThFeAsN agrees very well with other optimally doped compounds (red stars in Fig.4). Again, this results suggest ThFeAsN is nearly in optimized superconducting state with lots of itinerant electrons and away from the "parent" compound. Another fact should be noticed that is, the $h_{Fe-As}$ of ThFeAsN (1.2964 \AA) is lower than LaFeAsO (1.3166 \AA) \cite{CruzC,HuangQ} and SrFeAsF (1.3710 \AA) \cite{Hosono2,ZhuX2}. The closer distance of Fe-As will surely favor the electron hopping, thus reduce the electron correlations and orbital order controlled by the Hund¡¯s coupling $J_H$ within one atomic site \cite{TamY,HuD2}. This is a reasonable explanation for the absence of magnetic order, structural transition, and resistivity anomaly in ThFeAsN.

\section{Summary}
In summary, we have carried out neutron diffraction experiments on synthesized ThFeAsN over a temperature range of 6 K to 300 K. It is seen that there is neither structural transition nor magnetic transition existing in ThFeAsN, but a structural distortion may occur around 160 K. By comparing with other iron pnictides, we conclude that ThFeAsN may place near the optimal doping level and have reduced electron correlations.

\acknowledgments
The authors are grateful for the help on neutron scattering experiment from Dr. Guochu Deng at Australian Centre for Neutron Scattering, ANSTO.
This work is supported by the National Natural Science Foundation of China (Nos. 11374011, 11374346, 11674406, 11674372 and 11304183), the Strategic Priority Research
Program (B) of CAS (XDB07020300), the Ministry of Science and Technology of China (No. 2016YFA0300502) and the Youth Innovation Promotion Association of CAS (No. 2016004).
S. Li and H. Luo acknowledge the project supported by NPL, CAEP (No. 2015AB03).
\\

$^{\dag}$ Email: hqluo@iphy.ac.cn

\end{document}